\documentclass[prl,aps,twocolumn,longbibliography,10pt]{revtex4-2}

\pdfoutput=1

\date{\today}

\newcommand{\kB}{k_{\rm{B}}\!}
\newcommand{\Tc}{T_{\rm{c}}}
\newcommand{\um}{\upmu\textrm{m}}
\newcommand{\Vi}{V}
\newcommand{\Vf}{V_{\rm{f}}}
\newcommand{\Ti}{T}
\newcommand{\Tf}{T_{\rm{f}}}
\newcommand{\Tci}{T_{\rm{c}}}
\newcommand{\Tcf}{T_{{\rm{c},\rm{f}}}}
\newcommand{\Ek}{E_{\rm k}}
\newcommand{\Eint}{E_{\rm int}}

\usepackage{float}

\usepackage{fourier}

\usepackage{color}
\usepackage{graphicx}
\usepackage{amsmath,amssymb}

\usepackage{times}
\usepackage{upgreek}
\usepackage{psfrag} 
\usepackage{latexsym} 
\usepackage{amstext}
\usepackage{amsxtra} 
\usepackage{mathtools}

\usepackage{textcomp}
\usepackage{amsfonts}
\usepackage{graphicx}
\usepackage{bm}
\usepackage{color}
\usepackage{braket}
\usepackage{units}
\usepackage[svgnames]{xcolor}
\usepackage{soul}

\usepackage{silence}
\definecolor{myColor}{rgb}{0.02,0.12,0.3}
\definecolor{myciteColor}{rgb}{0.39,0.7,0.89}
\usepackage[colorlinks=true,citecolor=myColor,linkcolor=myColor,urlcolor=myColor]{hyperref}

\def\be{\begin{equation}}
\def\ee{\end{equation}}

\makeatletter
\def\@fnsymbol#1{\ensuremath{\ifcase#1\or \dagger\or *\or \ddagger\or
   \mathsection\or \mathparagraph\or \|\or **\or \dagger\dagger
   \or \ddagger\ddagger \else\@ctrerr\fi}}
\makeatother

\begin{document} 

\title{
Joule expansion of a quantum gas
}
\author{Christopher J.~Ho$^{*,\dag}$}
\author{Simon M.~Fischer$^*$}
\author{Gevorg~Martirosyan}
\author{Sebastian~J.~Morris}
\author{Ji\v r\' i~Etrych}
\author{Christoph~Eigen}
\author{Zoran~Hadzibabic}
\affiliation{Cavendish Laboratory, University of Cambridge, J. J. Thomson Avenue, Cambridge CB3 0HE, United Kingdom}

\begin{abstract}
We revisit the classic Joule-expansion experiments, now with a quantum-degenerate atomic Bose gas. In contrast to the classical-gas experiments, where no temperature change was measured, here we observe and quantitatively explain both cooling and heating effects, which arise, respectively, due to quantum statistics and inter-particle interactions.
\end{abstract}

\maketitle 

The Joule expansion of a gas, first performed by Gay-Lussac in 1807~\cite{GayLussac:1807} and then independently by Joule in 1845~\cite{Joule:1845}, is a classic experiment in thermodynamics that was crucial for proving the equivalence of work and heat, and establishing the first law of thermodynamics~\cite{Cardwell:1971}. In these experiments, a gas is initially confined to a part of a thermally insulated vessel and then allowed to freely expand and fill the entire volume without doing work on its surroundings. While the original experiments were celebrated for measuring no change in the gas temperature, it was later understood that this expansion is isothermal only for ideal (non-interacting) gases, and should lead to cooling or heating for gases with, respectively, attractive or repulsive interactions~\cite{Goussard:1993,Zemansky:1997}. Subsequent experiments sought to measure this interaction effect, but were unable to do so; in classical gases the expected fractional temperature change, set by the ratio of interaction to kinetic energy, is small~\footnote{Typically, this ratio is $\sim 10^{-3}$.}, and easily masked by any leakage of heat into or from the walls of the vessel~\cite{Roebuck:1930,Zemansky:1997}.

\begin{figure}[b]
\centerline{\includegraphics[width=\columnwidth]{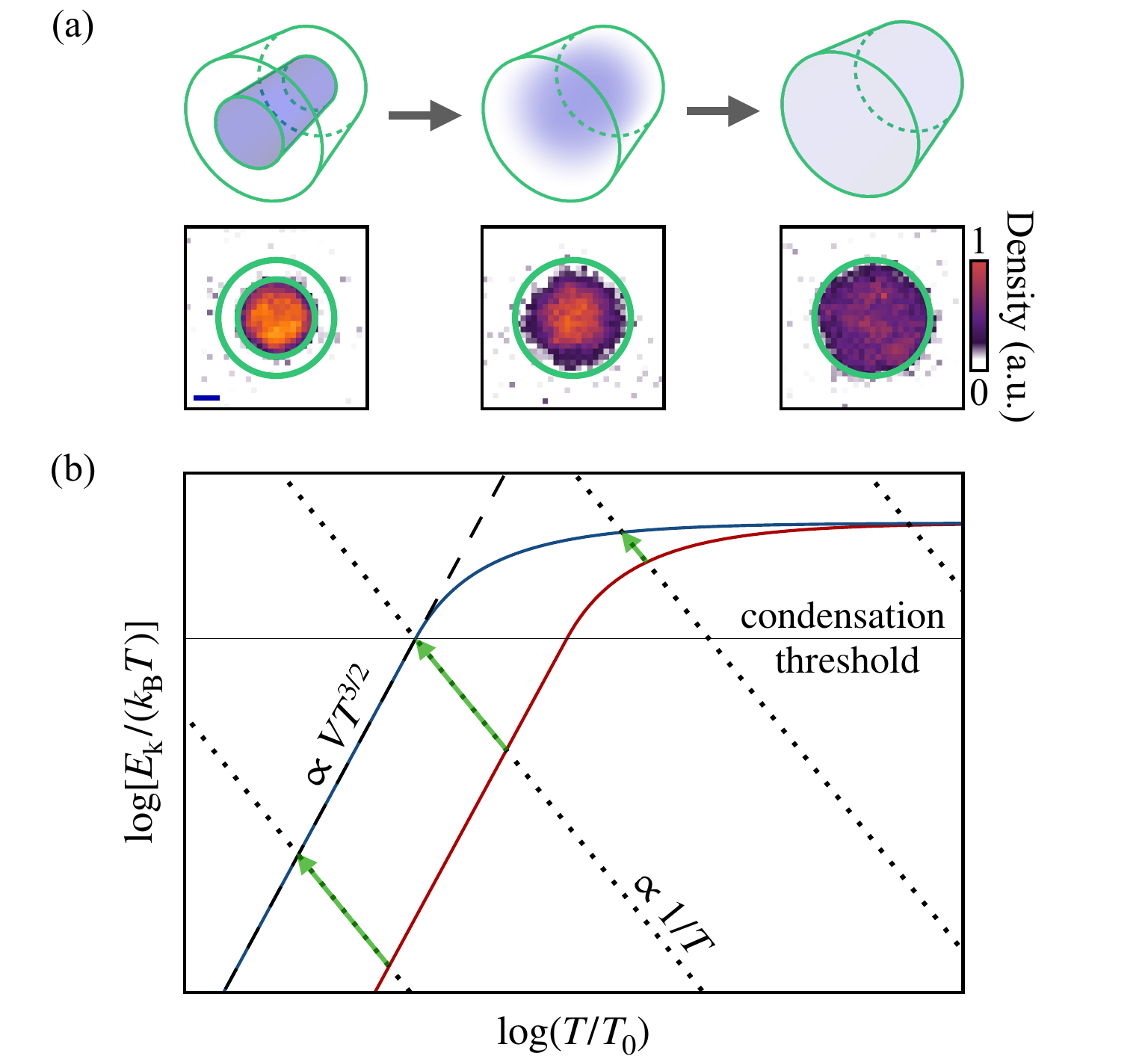}}
\caption{
Joule expansion of a Bose gas.
(a) Our setup consists of two cylindrical optical box traps, shown in green. We prepare a quantum-degenerate gas in the inner box, and, by switching off this trap, allow the gas to freely expand and fill the outer one.  The bottom row shows absorption images of the clouds before, during, and after the expansion. The scale bar (blue) corresponds to $10\,\um$.
(b)~Thermodynamic diagram for the ideal Bose gas.
Here $\Ek$ is the kinetic energy per particle, $T$ is the temperature, and $T_0$ is an arbitrary constant introduced just to define the dimensionless temperature $T/T_0$.
The red and blue solid curves are isochors for the same atom number $N$ but different volumes $V$, so different condensation temperatures.
Below the condensation temperature $\Ek \propto VT^{5/2}$ (dashed line), while in the classical-gas limit $\Ek=(3/2)\, \kB T$. The dotted lines are isoenergetic contours, and the green arrows show the Joule expansion for various initial temperatures in the smaller volume (red).
}
\label{fig1}
\vspace{-0.5em}
\end{figure}

In this Letter, we revisit Joule-expansion experiments with a quantum-degenerate atomic Bose gas, in which the interaction energy can exceed the kinetic one, and for which a purely quantum-statistical isoenergetic cooling is expected in the absence of interactions. Our setup consists of two concentric optical box traps~\cite{Gaunt:2013,Navon:2021}, as outlined in Fig.~\ref{fig1}(a), and the strength of contact interactions in our \textsuperscript{39}K gas is characterized by the $s$-wave scattering length $a$.
Previously, the related quantum Joule--Thomson (JT) effect~\cite{Kothari:1937} was observed in a single optical box~\cite{Schmidutz:2014, Ji:2024}, by mapping the JT isoenthalpic rarefaction onto removal of particles from the trap, but this mapping is valid only in the special cases $a\rightarrow 0$ and $a\rightarrow \infty$~(see also~\footnote{In harmonically trapped quantum gases many fascinating thermodynamic studies were performed in the regime where the interaction energy is comparable to or dominates the kinetic one~\cite{Gerbier:2004, Nascimbene:2010,Navon:2011,Yefsah:2011,Ku:2012,Mordini:2020}, but these systems are not naturally suited for the textbook gas-expansion experiments.}).
Our setup allows studies of Joule expansion for any interaction strength, and by tuning $a$ via a magnetic Feshbach resonance, we observe both  isoenergetic cooling for vanishing interactions ($a\rightarrow 0$) and isoenergetic heating due to repulsive interactions ($a>0$). 

Our experiments begin with a quasi-homogeneous gas of \textsuperscript{39}K atoms held in the inner box trap. We prepare the gas in the lowest hyperfine state and control the scattering length $a$ by tuning the magnetic field $B$ in the vicinity of the resonance at $402.7\,\textrm{G}$~\cite{Etrych:2023}. We then rapidly switch off the inner trap, allowing the gas to fill the outer one, and measure the temperature after the gas has thermalized~\cite{Supplementary}.

We start with the ideal-gas case (\mbox{$a\rightarrow~0$}) and in Fig.~\ref{fig1}(b) graphically explain the origin of the purely quantum-statistical cooling of a homogeneous Bose gas. For the same atom number $N$ and two different volumes $V$, we plot the dimensionless $\Ek/(\kB T)$ versus the dimensionless $T/T_0$ (solid lines), where $\Ek$ is the kinetic energy per particle (here equal to the total energy per particle), $T$ is the temperature, and $T_0$ an arbitrary reference.
The dotted lines are isoenergetic contours, and the green arrows show the cooling effect of Joule expansion from the smaller (red) to the larger (blue) volume, which vanishes in the high-$T$ limit; the ratios of final to initial temperatures are given by the horizontal components of the green vectors.

Quantitatively:
\be
\dfrac{\Ek}{\kB T} =
\begin{dcases*}
\dfrac{3}{2}
\dfrac{\zeta(5/2)}{\zeta(3/2)} \left(\dfrac{T}{\Tc}\right)^{3/2} & for $T \leq \Tc$ \\
\dfrac{3}{2}
\dfrac{g_{5/2}(z)}{g_{3/2}(z)} & for $T > \Tc$
\end{dcases*}\,,
\label{eq:chi}
\ee
where $\Tc \propto (N/V)^{2/3}$ is the critical temperature for Bose--Einstein condensation, $\zeta$ is the Riemann function, with $\zeta(5/2)/\zeta(3/2) \approx 0.51$, $g_\alpha$ is a polylogarithm of order $\alpha$, and $z = \exp[{\mu/(\kB T)}]$ is the fugacity, where $\mu$ is the chemical potential. For $T>\Tc$, the fugacity satisfies 
\be
g_{3/2}(z) = \frac{N\lambda^3}{V}\,,
\ee
where $\lambda = h / \sqrt{2\pi m\kB T}$ is the thermal wavelength and $m$ the atom mass.
Below $\Tc$, Eq.~(\ref{eq:chi}) gives $\Ek/(\kB T) \propto VT^{3/2}$ [dashed line in Fig.~\ref{fig1}(b)], while above $\Tc$ the dependence on $V$ and $T$ is implicit in $z$. For the classical gas $z\ll1$; in this limit $g_{\alpha}(z) \rightarrow z$ and $\Ek = (3/2)\kB T$ is independent of $V$, so the cooling effect vanishes.

From hereon, we denote the initial volume and temperature (in the inner box) $\Vi$ and $\Ti$, and the final volume and temperature (in the outer box) $\Vf$ and $\Tf$. Denoting the critical temperature in the inner box $\Tc$, the critical temperature in the outer one is $\Tcf = \Tc(V/\Vf)^{2/3}$.

If the gas is condensed both before and after the expansion, Eq.~(\ref{eq:chi}) gives
\be
\frac{\Tf}{\Ti} = \left(\frac{\Vi}{\Vf}\right)^{2/5}.
\label{eq:jouleScaling}
\ee
This relation holds if $\Tf \leq \Tcf$, which, using Eq.~(\ref{eq:jouleScaling}), gives the condition for the initial temperature
\be
\frac{\Ti}{\Tci} \leq \left(\frac{\Vi}{\Vf}\right)^{4/15}.
\label{eq:jouleCondition}
\ee
For higher $\Ti/\Tc$, the fractional cooling, $1-\Tf/\Ti$, gradually reduces and vanishes in the classical-gas limit.

Experimentally, we prepare an equilibrium interacting gas of density $n\approx 1.7\,\um^{-3}$ in the inner box, with $\Vi \approx 3.6\times10^4\,\um^3$, and tune $a\rightarrow 0$ before the expansion. After the gas fills the outer box, with $\Vf \approx 2.6\, \Vi$, we increase the scattering length to $a=300\,a_0$ (where $a_0$ is the Bohr radius) to facilitate thermalization (see also~\cite{Supplementary}).

\begin{figure}[t]
\centerline{\includegraphics[width=\columnwidth]{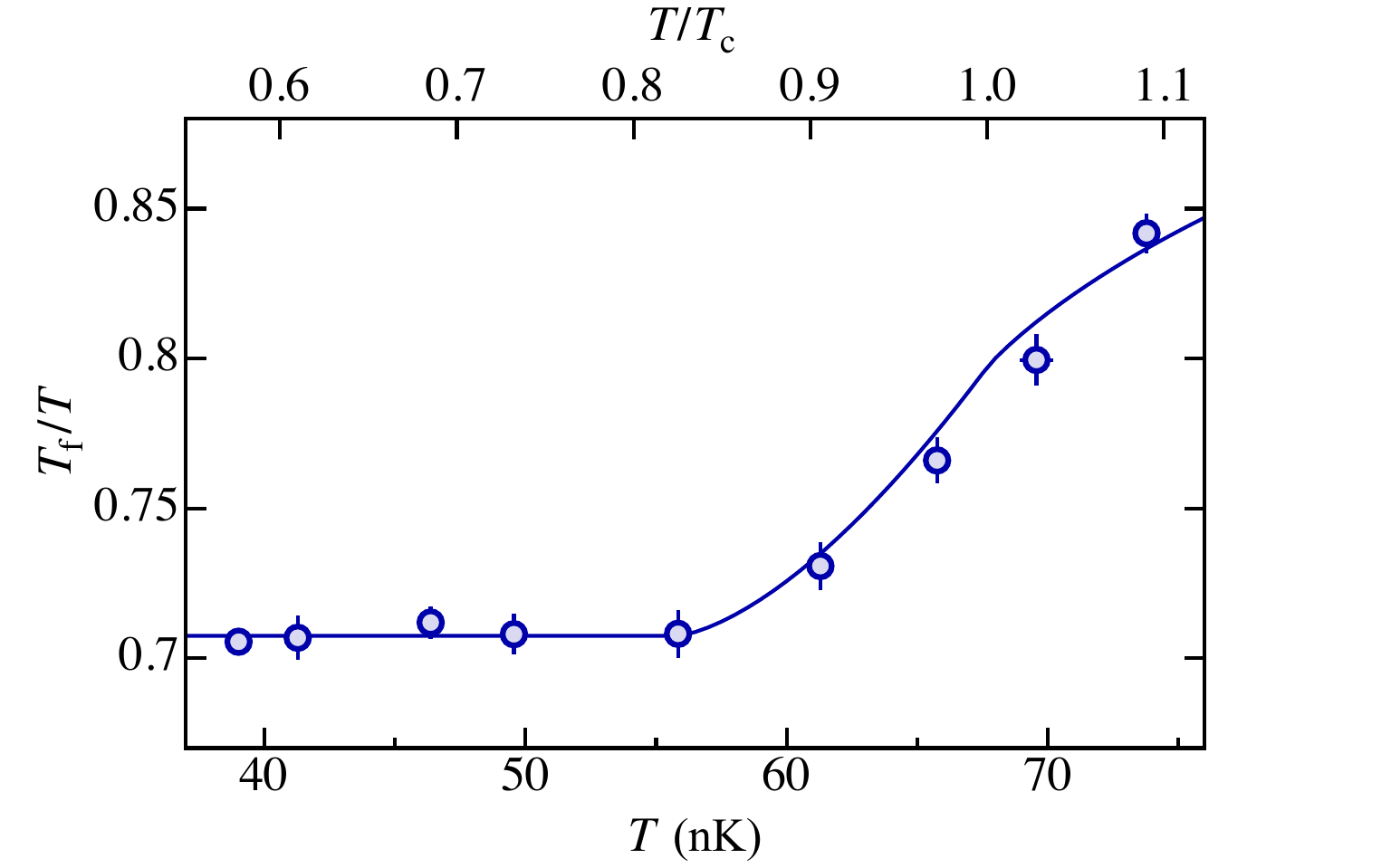}}
\caption{
Quantum-statistical cooling of a non-interacting Bose gas due to Joule expansion. Here $\Ti$ is the initial temperature, $\Tc$ the condensation temperature for the initial gas density, and $\Tf$ the final temperature.
The fractional temperature change, $1 - \Tf/\Ti$, is generally greater for lower $\Ti/\Tci$, and constant if the gas is partially condensed both before and after the expansion. Our measurements agree with numerical calculations (solid line) that take into account the fact that the gas in our optical box trap is not perfectly homogeneous~\cite{Supplementary}; in a perfect box, the low-$T$ plateau of $\Tf/T$ would, as per Eqs.~(\ref{eq:jouleScaling}) and~(\ref{eq:jouleCondition}), have a value of $0.68$ and extend to $T/\Tc \approx 0.78$.
}
\label{fig2}
\vspace{-1em}
\end{figure}

In Fig.~\ref{fig2}, we plot the measured $\Tf/\Ti$ versus both $\Ti$ and $\Ti/\Tci$. Our observations agree with numerical calculations (solid line) that take into account the fact that the gas in our optical box is not perfectly homogeneous~\cite{Supplementary}. For comparison, the low-$\Ti/\Tci$ plateau is at $\Tf/\Ti \approx 0.71$, whereas according to Eq.~(\ref{eq:jouleScaling}) it would be at $\Tf/T\approx 0.68$, and the plateau extends to $T/\Tc \approx 0.83$, whereas according to Eq.~(\ref{eq:jouleCondition}) it would extend to $T/\Tc \approx 0.78$.

We now turn to experiments with an interacting quantum gas, where the total energy per particle, conserved during the Joule expansion, is 
\be
E = \Ek + \Eint \, ,
\label{eq:Etot}
\ee
where $\Eint$ is the interaction energy per particle.
In the Hartree--Fock approximation~\cite{Pethick:2002}, 
\be
\Eint = (2-\eta^2) \, \frac{2\pi\hbar^2na}{m}\, ,
\label{eq:EintHF}
\ee
where $\eta$ is the condensed fraction; the factor $2-\eta^2$ quantifies two-body correlations in the quantum gas. Relative to the ideal-gas case, the presence of repulsive interactions ($a>0)$ during the isoenergetic expansion raises $\Tf$, because, as the density drops, $\Eint$ is partially converted into $\Ek$.

\begin{figure*}[t]
\centerline{\includegraphics[width=\textwidth]{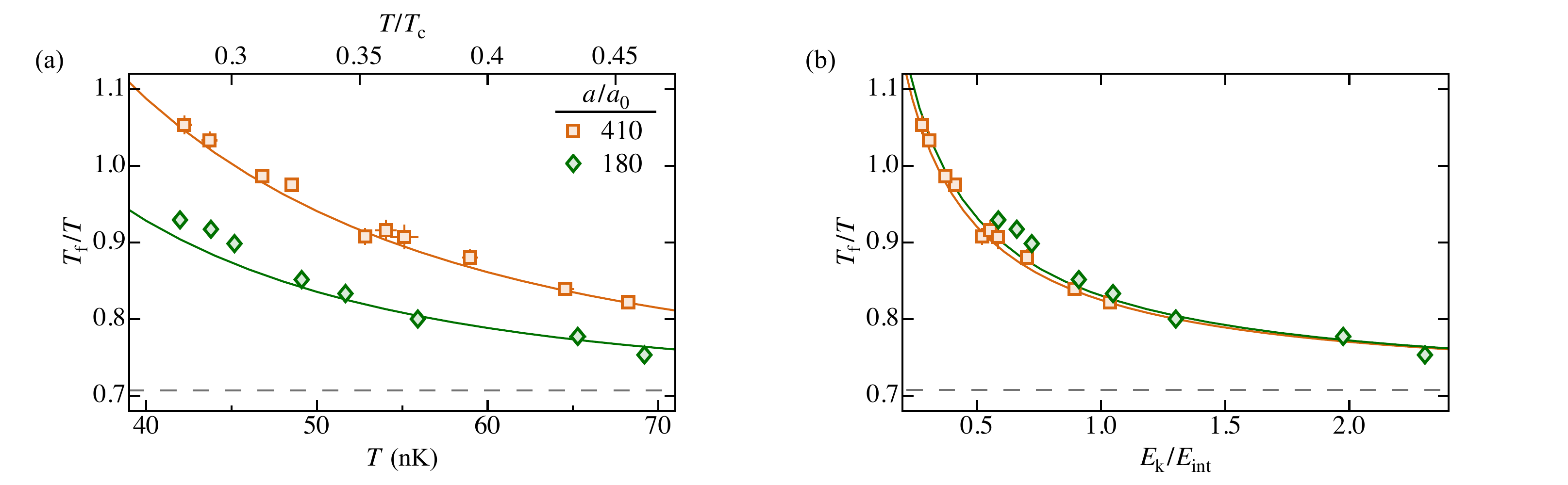}}
\caption{
Joule expansion of an interacting Bose gas.
(a) The quantum-statistical cooling effect, indicated by the dashed line, is countered by a heating effect due to repulsive interactions.
Our measurements for two values of the scattering length $a$ are reproduced in numerical calculations~\cite{Supplementary}, shown by the solid lines.
(b) Plotting the same data versus the ratio of the initial kinetic and interaction energies shows that the interaction effect overcomes the quantum-statistical one (so $\Tf > \Ti$) for $\Eint/\Ek \gtrsim 3$.
}
\label{fig3}
\vspace{-1em}
\end{figure*}

In experiments, we now increase the initial density to $n \approx 7\,\um^{-3}$, and focus on the low-$\Ti/\Tc$ regime, where the purely quantum-statistical cooling effect is constant. In Fig.~\ref{fig3}(a), we show measurements of $\Tf/\Ti$ for two different values of $a$, which agree with our numerical calculations~\cite{Supplementary}. In Fig.~\ref{fig3}(b) we plot the same data versus the initial $\Ek/\Eint$, which shows that, for our parameters, the interaction effect overcomes the quantum-statistical one (so $\Tf > \Ti$) for $\Eint/\Ek \gtrsim 3$. Note that the two (theoretical) curves do not perfectly coincide because for the same initial $\eta$ the change in two-body correlations due to the expansion depends on $a$.

In conclusion, we have revisited the classic Joule-expansion experiments with a quantum Bose gas, and quantitatively revealed the richer physics that arises through an interplay of quantum statistics, inter-particle interactions, and quantum correlations. With current technology~\cite{Navon:2021, Gauthier:2021}, our methods could also be extended to slow (adiabatic) expansions, which would allow realizations of closed thermodynamic cycles and quantum engines in box-like geometries~\cite{Myers:2022,Koch:2023,Estrada:2024}.

The data that support the findings of this article
are openly available~\cite{repository}. 

This work was supported by ERC [UniFlat] and STFC [Grants No.~ST/T006056/1 and No.~ST/Y004469/1]. Z.H. acknowledges support from the Royal Society Wolfson Fellowship.

\newpage
\cleardoublepage

\setcounter{figure}{0} 
\setcounter{equation}{0} 

\renewcommand\theequation{S\arabic{equation}} 
\renewcommand\thefigure{S\arabic{figure}} 

\section{Supplemental Material}

\subsection{\textsc{Calculations for imperfect box trap}}

The walls of optical box traps~\cite{Navon:2021}, created by focused laser beams, are not infinitely steep, and the resulting deviation from a perfect box potential is, for the purposes of thermodynamic calculations, captured well by an isotropic power-law potential, $U(r) = U_0 (r/r_0)^p$, where $p\gg 1$ and $r_0$ defines the effective box size. In our setup~\cite{Eigen:2016}, we typically achieve $p\approx 20$ for low-temperature ($T\ll \Tc$) gases, but to study clouds at temperatures approaching $\Tc$ we increase the trapping-laser power, which reduces the exponent $p$. We experimentally assess that $p \approx 10$ by studying the shape of the momentum distribution in a degenerate non-condensed gas~\cite{Gaunt:2013,Schmidutz:2014}.

In the semi-classical approximation, the distribution function for the ideal Bose gas in such a potential is
\be
f_{\rm B}(k,r) = \frac{1}{(2\pi)^3} \, g_0\left( z e^{- \frac{\hbar^2k^2}{2m\kB T} - \frac{U(r)}{\kB T}} \right) \, ,
\ee
and integrating over $r$ gives the momentum distribution:
\be
f_{\textrm{B}}(k) = \frac{V}{(2\pi)^3} \, g_{3/p} \left( ze^{-\frac{\hbar^2k^2}{2m\kB T}} \right)\, ,
\label{eq:nk_3D_n}
\ee
where
\be
V \equiv \frac{3}{p}\, \Gamma\left(\frac{3}{p}\right)\, \left(\frac{\kB T}{U_0}\right)^{3/p}\, \frac{4\pi}{3}r_0^3 \, ,
\label{eq:Volume}
\ee
and $\Gamma$ is the Gamma function; note that $V\rightarrow (4\pi/3) \, r_0^3$ for $p\rightarrow\infty$. Integrating $f_{\textrm{B}}(k)$ over $k$ gives the total number of non-condensed atoms: 
\be
N' = \frac{V}{\lambda^3} \, g_{\alpha}(z)\, ,
\label{eq:NImperfect}
\ee
with $\alpha = 3/2+3/p$. Also note that $V/\lambda^3 \propto T^{\alpha}$. The critical temperature for condensation is then given by $z=0$ and $N' = N$, where $N$ is the total particle number, so $\Tc \propto N^{1/\alpha}$. 

Similarly, starting from Eq.~(\ref{eq:nk_3D_n}) and integrating for the kinetic energy per particle:
\be
\dfrac{\Ek}{\kB T} = 
\begin{dcases*}
\frac{3}{2} 
\frac{\zeta(\alpha + 1)}{\zeta(\alpha)} \left(\frac{T}{\Tc}\right)^\alpha & for $T \leq \Tc$ \\
\frac{3}{2} 
\frac{g_{\alpha+1}(z)}{g_{\alpha}(z)} & for $T > \Tc$
\end{dcases*}
\, .
\label{eq:EImperfect}
\ee

\begin{figure}[t!]
\centerline{
\includegraphics[width=\columnwidth]{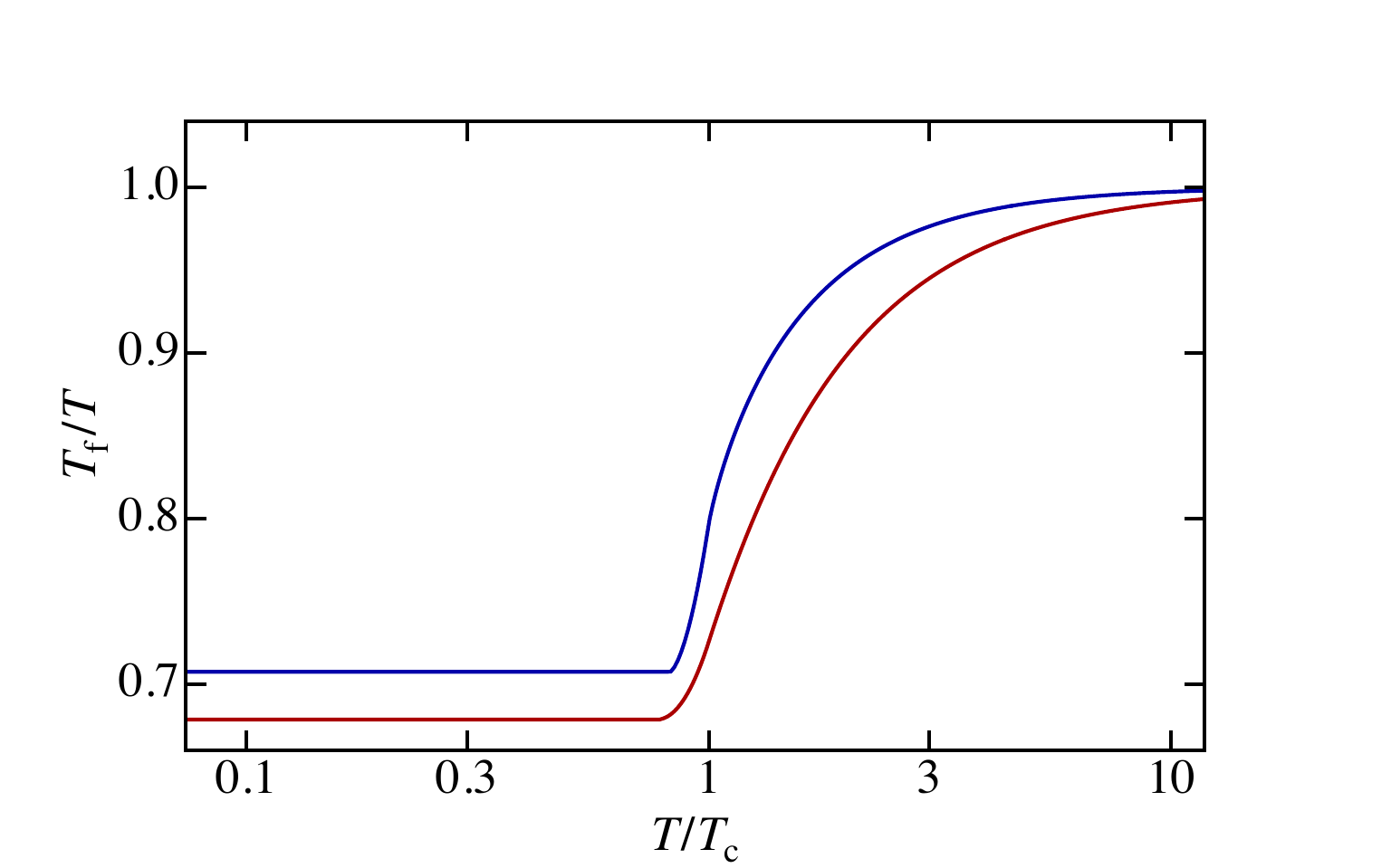}}
\caption{Ideal-gas calculations of $\Tf/\Ti$ as a function of $\Ti/\Tc$, for both $p=10$ (blue) and $p\rightarrow \infty$ (red).
}
\label{figS1}
\end{figure}

If the gas is partially condensed both before and after the expansion,
\be
\frac{\Tf}{\Ti} = \left(\frac{\Vi}{\Vf}\right)^{\frac{1}{\alpha+1}},
\label{eq:jouleCoolingImperfectBox}
\ee
and more generally we numerically solve Eqs.~(\ref{eq:NImperfect}) and (\ref{eq:EImperfect}) to get $\Tf$.
In Fig.~\ref{figS1} we show the results of our calculations for a larger range of $\Ti/\Tc$ than covered in the experiments, for both $p=10$ and a perfect box potential ($p\rightarrow \infty$).

For the interacting gas, to calculate $\Eint$ we define $n=N/V$, self-consistently taking into account that $V \propto T^{3/p}$.

\vspace{1em}
\subsection{\textsc{Temperature measurements}}

To measure the temperature in either the inner or the outer box, we switch off interactions ($a \rightarrow 0$), release the clouds, and image them along the box axis after $16-20$\,ms of time-of-flight expansion.
This gives the line-of-sight integrated momentum distributions, corresponding to the distribution in Eq.~(\ref{eq:nk_3D_n}) integrated along one direction.

\begin{figure*}[t]
\centerline{
\includegraphics[width=\textwidth]{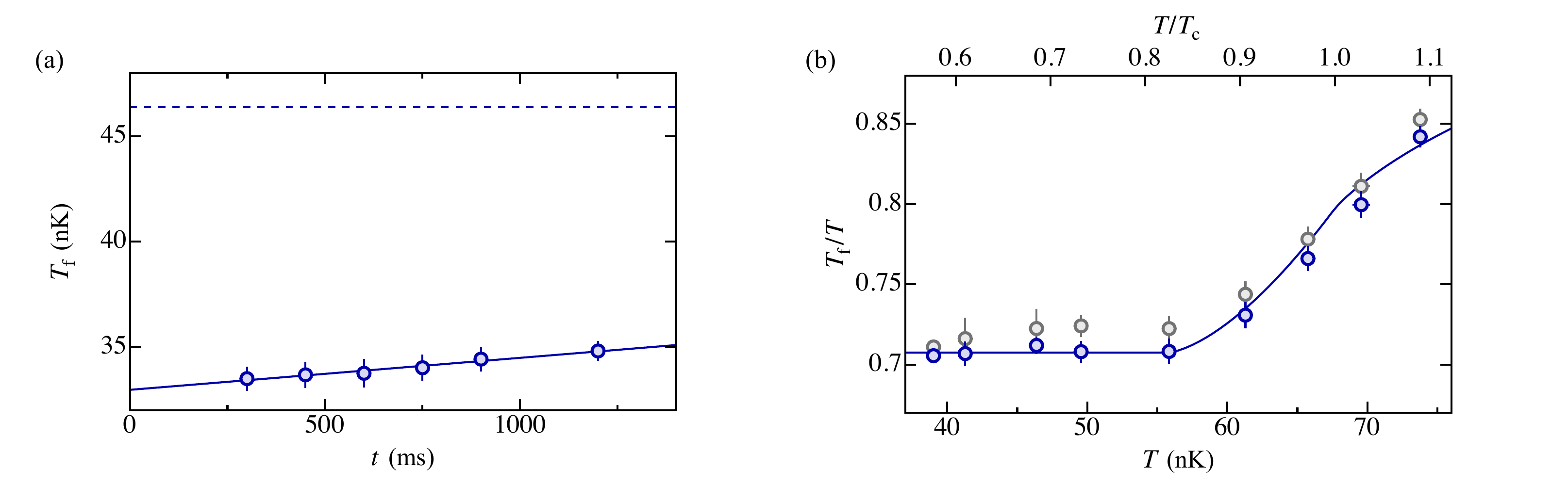}}
\caption{
Measurements of $\Tf$ in the ideal-gas experiments. (a) Here $t=0$ corresponds to the time when the thermalizing interactions have been fully turned on (see text). The thermalization time is $\lesssim 300\,$ms, but we measure $\Tf$ for $t$ values up to $1.2\,$s in order to account for the slow technical heating, here at a rate of $1.6\,{\rm nK}/{\rm s}$ (slope of the solid line). The $\Tf$ values reported in the paper are obtained by linear extrapolation to $t=0$, which corrects $\Tf/\Ti$ values at a $\lesssim 0.01$ level. The dashed line shows the temperature before the Joule expansion.
(b) Comparison of the deduced $\Tf/\Ti$ with (blue symbols) and without (gray symbols) accounting for the slow technical heating; the blue-symbols data is the same as in Fig.~2 in the main paper.
}
\label{figS2}
\end{figure*}

We measure the initial temperature $T$ right before the inner box is turned off, while the final temperature $\Tf$ is defined only after the gas thermalizes in the outer box. In our ideal-gas experiments, with $a$ ramped to $300\,a_0$ for the thermalization, the thermalization time [$\propto 1/(n_{\rm f}\, a^2)$, where $n_{\rm f}$ is the final density] is $\lesssim 300\,$ms, while the technical heating rate in our outer trap is about $2\,{\rm nK}/{\rm s}$ at $300\,a_0$ [see Fig.~\ref{figS2}(a)], and lower for smaller $a$. This heating can affect the deduced $\Tf$ at a $0.5\,$nK level.
As we illustrate in Fig.~\ref{figS2}(a), to account for this effect, we measure $\Tf$ over much longer hold times in the outer trap; note that we ramp $a$ to $300\,a_0$ over $100\,$ms and here $t=0$ corresponds to the end of that ramp (we have also checked that changing the ramp time to $300\,$ms or $500\,$ms does not change our results within errors).
Then, as the best estimate of the true $\Tf$ (for instantaneous thermalization or in absence of technical heating), we extrapolate the measured temperature linearly to $t=0$, as shown by the solid line. Compared to simply measuring $\Tf$ at $t=300\,$ms, this changes our $\Tf/\Ti$ values by $\lesssim 0.01$; see Fig.~\ref{figS2}(b). We apply the same method for the interacting-gas measurements in Fig.~3, where the heating effect is similar in size.


\begin{thebibliography}{26}%
\makeatletter
\providecommand \@ifxundefined [1]{%
 \@ifx{#1\undefined}
}%
\providecommand \@ifnum [1]{%
 \ifnum #1\expandafter \@firstoftwo
 \else \expandafter \@secondoftwo
 \fi
}%
\providecommand \@ifx [1]{%
 \ifx #1\expandafter \@firstoftwo
 \else \expandafter \@secondoftwo
 \fi
}%
\providecommand \natexlab [1]{#1}%
\providecommand \enquote  [1]{``#1''}%
\providecommand \bibnamefont  [1]{#1}%
\providecommand \bibfnamefont [1]{#1}%
\providecommand \citenamefont [1]{#1}%
\providecommand \href@noop [0]{\@secondoftwo}%
\providecommand \href [0]{\begingroup \@sanitize@url \@href}%
\providecommand \@href[1]{\@@startlink{#1}\@@href}%
\providecommand \@@href[1]{\endgroup#1\@@endlink}%
\providecommand \@sanitize@url [0]{\catcode `\\12\catcode `\$12\catcode `\&12\catcode `\#12\catcode `\^12\catcode `\_12\catcode `\%12\relax}%
\providecommand \@@startlink[1]{}%
\providecommand \@@endlink[0]{}%
\providecommand \url  [0]{\begingroup\@sanitize@url \@url }%
\providecommand \@url [1]{\endgroup\@href {#1}{\urlprefix }}%
\providecommand \urlprefix  [0]{URL }%
\providecommand \Eprint [0]{\href }%
\providecommand \doibase [0]{https://doi.org/}%
\providecommand \selectlanguage [0]{\@gobble}%
\providecommand \bibinfo  [0]{\@secondoftwo}%
\providecommand \bibfield  [0]{\@secondoftwo}%
\providecommand \translation [1]{[#1]}%
\providecommand \BibitemOpen [0]{}%
\providecommand \bibitemStop [0]{}%
\providecommand \bibitemNoStop [0]{.\EOS\space}%
\providecommand \EOS [0]{\spacefactor3000\relax}%
\providecommand \BibitemShut  [1]{\csname bibitem#1\endcsname}%
\let\auto@bib@innerbib\@empty
\item[$^{\color{myColor}\ast}$] These authors contributed equally to this work.
\item[$^{\color{myColor}\dag}$] Current address: Centre for Cold Matter, Blackett Laboratory, Imperial College London, Prince Consort Road, London SW7 2AZ, United Kingdom; Email: \href{mailto:cjh211@ic.ac.uk}{cjh211@ic.ac.uk}
\bibitem [{\citenamefont {Gay-Lussac}(1807)}]{GayLussac:1807}%
  \BibitemOpen
  \bibfield  {author} {\bibinfo {author} {\bibfnamefont {L.~J.}\ \bibnamefont {Gay-Lussac}},\ }\bibfield  {title} {\bibinfo {title} {First attempt to determine the changes in temperature which gases experience owing to changes of density, and considerations on their capacity of heat},\ }\href@noop {} {\bibfield  {journal} {\bibinfo  {journal} {Mem. Soc. d'Arcueil}\ }\textbf {\bibinfo {volume} {1}},\ \bibinfo {pages} {3} (\bibinfo {year} {1807})}\BibitemShut {NoStop}%
\bibitem [{\citenamefont {Joule}(1845)}]{Joule:1845}%
  \BibitemOpen
  \bibfield  {author} {\bibinfo {author} {\bibfnamefont {J.~P.}\ \bibnamefont {Joule}},\ }\bibfield  {title} {\bibinfo {title} {On the changes of temperature produced by the rarefaction and condensation of air},\ }\href {https://doi.org/10.1080/14786444508645153} {\bibfield  {journal} {\bibinfo  {journal} {Lond. Edinb. Dubl. Phil. Mag.}\ }\bibinfo {series} {S3},\ \textbf {\bibinfo {volume} {26}},\ \bibinfo {pages} {369} (\bibinfo {year} {1845})}\BibitemShut {NoStop}%
\bibitem [{\citenamefont {Cardwell}(1971)}]{Cardwell:1971}%
  \BibitemOpen
  \bibfield  {author} {\bibinfo {author} {\bibfnamefont {D.~S.~L.}\ \bibnamefont {Cardwell}},\ }\href@noop {} {\emph {\bibinfo {title} {From Watt to Clausius: The rise of thermodynamics in the early industrial age}}}\ (\bibinfo  {publisher} {Cornell University Press},\ \bibinfo {year} {1971})\BibitemShut {NoStop}%
\bibitem [{\citenamefont {Goussard}\ and\ \citenamefont {Roulet}(1993)}]{Goussard:1993}%
  \BibitemOpen
  \bibfield  {author} {\bibinfo {author} {\bibfnamefont {J.-O.}\ \bibnamefont {Goussard}}\ and\ \bibinfo {author} {\bibfnamefont {B.}~\bibnamefont {Roulet}},\ }\bibfield  {title} {\bibinfo {title} {Free expansion for real gases},\ }\href {https://doi.org/10.1119/1.17417} {\bibfield  {journal} {\bibinfo  {journal} {Am. J. Phys.}\ }\textbf {\bibinfo {volume} {61}},\ \bibinfo {pages} {845} (\bibinfo {year} {1993})}\BibitemShut {NoStop}%
\bibitem [{\citenamefont {Zemansky}\ and\ \citenamefont {Dittman}(1997)}]{Zemansky:1997}%
  \BibitemOpen
  \bibfield  {author} {\bibinfo {author} {\bibfnamefont {M.~W.}\ \bibnamefont {Zemansky}}\ and\ \bibinfo {author} {\bibfnamefont {R.~H.}\ \bibnamefont {Dittman}},\ }\href@noop {} {\emph {\bibinfo {title} {Heat and Thermodynamics}}},\ \bibinfo {edition} {7th}\ ed.\ (\bibinfo  {publisher} {McGraw-Hill},\ \bibinfo {year} {1997})\BibitemShut {NoStop}%
\bibitem [{Note1()}]{Note1}%
  \BibitemOpen
  \bibinfo {note} {Typically, this ratio is $\sim 10^{-3}$.}\BibitemShut {Stop}%
\bibitem [{\citenamefont {Roebuck}(1930)}]{Roebuck:1930}%
  \BibitemOpen
  \bibfield  {author} {\bibinfo {author} {\bibfnamefont {J.~R.}\ \bibnamefont {Roebuck}},\ }\bibfield  {title} {\bibinfo {title} {{The {Joule-Thomson} Effect in Air. Second Paper}},\ }\href {https://doi.org/10.2307/20026275} {\bibfield  {journal} {\bibinfo  {journal} {Proc. Am. Acad. Arts Sci.}\ }\textbf {\bibinfo {volume} {64}},\ \bibinfo {pages} {287} (\bibinfo {year} {1930})}\BibitemShut {NoStop}%
\bibitem [{\citenamefont {Gaunt}\ \emph {et~al.}(2013)\citenamefont {Gaunt}, \citenamefont {Schmidutz}, \citenamefont {Gotlibovych}, \citenamefont {Smith},\ and\ \citenamefont {Hadzibabic}}]{Gaunt:2013}%
  \BibitemOpen
  \bibfield  {author} {\bibinfo {author} {\bibfnamefont {A.~L.}\ \bibnamefont {Gaunt}}, \bibinfo {author} {\bibfnamefont {T.~F.}\ \bibnamefont {Schmidutz}}, \bibinfo {author} {\bibfnamefont {I.}~\bibnamefont {Gotlibovych}}, \bibinfo {author} {\bibfnamefont {R.~P.}\ \bibnamefont {Smith}},\ and\ \bibinfo {author} {\bibfnamefont {Z.}~\bibnamefont {Hadzibabic}},\ }\bibfield  {title} {\bibinfo {title} {{Bose--Einstein Condensation of Atoms in a Uniform Potential}},\ }\href {https://doi.org/10.1103/PhysRevLett.110.200406} {\bibfield  {journal} {\bibinfo  {journal} {Phys. Rev. Lett.}\ }\textbf {\bibinfo {volume} {110}},\ \bibinfo {pages} {200406} (\bibinfo {year} {2013})}\BibitemShut {NoStop}%
\bibitem [{\citenamefont {Navon}\ \emph {et~al.}(2021)\citenamefont {Navon}, \citenamefont {Smith},\ and\ \citenamefont {Hadzibabic}}]{Navon:2021}%
  \BibitemOpen
  \bibfield  {author} {\bibinfo {author} {\bibfnamefont {N.}~\bibnamefont {Navon}}, \bibinfo {author} {\bibfnamefont {R.~P.}\ \bibnamefont {Smith}},\ and\ \bibinfo {author} {\bibfnamefont {Z.}~\bibnamefont {Hadzibabic}},\ }\bibfield  {title} {\bibinfo {title} {Quantum gases in optical boxes},\ }\href {https://doi.org/10.1038/s41567-021-01403-z} {\bibfield  {journal} {\bibinfo  {journal} {Nat. Phys.}\ }\textbf {\bibinfo {volume} {17}},\ \bibinfo {pages} {1334} (\bibinfo {year} {2021})}\BibitemShut {NoStop}%
\bibitem [{\citenamefont {Kothari}\ and\ \citenamefont {Srinivasa}(1937)}]{Kothari:1937}%
  \BibitemOpen
  \bibfield  {author} {\bibinfo {author} {\bibfnamefont {D.}~\bibnamefont {Kothari}}\ and\ \bibinfo {author} {\bibfnamefont {B.}~\bibnamefont {Srinivasa}},\ }\bibfield  {title} {\bibinfo {title} {{Joule-Thomson effect and Quantum Statistics}},\ }\href {https://doi.org/10.1038/140970b0} {\bibfield  {journal} {\bibinfo  {journal} {Nature}\ }\textbf {\bibinfo {volume} {140}},\ \bibinfo {pages} {970} (\bibinfo {year} {1937})}\BibitemShut {NoStop}%
\bibitem [{\citenamefont {Schmidutz}\ \emph {et~al.}(2014)\citenamefont {Schmidutz}, \citenamefont {Gotlibovych}, \citenamefont {Gaunt}, \citenamefont {Smith}, \citenamefont {Navon},\ and\ \citenamefont {Hadzibabic}}]{Schmidutz:2014}%
  \BibitemOpen
  \bibfield  {author} {\bibinfo {author} {\bibfnamefont {T.~F.}\ \bibnamefont {Schmidutz}}, \bibinfo {author} {\bibfnamefont {I.}~\bibnamefont {Gotlibovych}}, \bibinfo {author} {\bibfnamefont {A.~L.}\ \bibnamefont {Gaunt}}, \bibinfo {author} {\bibfnamefont {R.~P.}\ \bibnamefont {Smith}}, \bibinfo {author} {\bibfnamefont {N.}~\bibnamefont {Navon}},\ and\ \bibinfo {author} {\bibfnamefont {Z.}~\bibnamefont {Hadzibabic}},\ }\bibfield  {title} {\bibinfo {title} {{Quantum Joule-Thomson Effect in a Saturated Homogeneous Bose Gas}},\ }\href {https://doi.org/10.1103/PhysRevLett.112.040403} {\bibfield  {journal} {\bibinfo  {journal} {Phys. Rev. Lett.}\ }\textbf {\bibinfo {volume} {112}},\ \bibinfo {pages} {040403} (\bibinfo {year} {2014})}\BibitemShut {NoStop}%
\bibitem [{\citenamefont {Ji}\ \emph {et~al.}(2024)\citenamefont {Ji}, \citenamefont {Chen}, \citenamefont {Schumacher}, \citenamefont {Assump\c{c}ao}, \citenamefont {Huang}, \citenamefont {Vivanco},\ and\ \citenamefont {Navon}}]{Ji:2024}%
  \BibitemOpen
  \bibfield  {author} {\bibinfo {author} {\bibfnamefont {Y.}~\bibnamefont {Ji}}, \bibinfo {author} {\bibfnamefont {J.}~\bibnamefont {Chen}}, \bibinfo {author} {\bibfnamefont {G.~L.}\ \bibnamefont {Schumacher}}, \bibinfo {author} {\bibfnamefont {G.~G.~T.}\ \bibnamefont {Assump\c{c}ao}}, \bibinfo {author} {\bibfnamefont {S.}~\bibnamefont {Huang}}, \bibinfo {author} {\bibfnamefont {F.~J.}\ \bibnamefont {Vivanco}},\ and\ \bibinfo {author} {\bibfnamefont {N.}~\bibnamefont {Navon}},\ }\bibfield  {title} {\bibinfo {title} {{Observation of the Fermionic Joule-Thomson Effect}},\ }\href {https://doi.org/10.1103/PhysRevLett.132.153402} {\bibfield  {journal} {\bibinfo  {journal} {Phys. Rev. Lett.}\ }\textbf {\bibinfo {volume} {132}},\ \bibinfo {pages} {153402} (\bibinfo {year} {2024})}\BibitemShut {NoStop}%
\bibitem [{Note2()}]{Note2}%
  \BibitemOpen
  \bibinfo {note} {In harmonically trapped quantum gases many fascinating thermodynamic studies were performed in the regime where the interaction energy is comparable to or dominates the kinetic one~\cite {Gerbier:2004, Nascimbene:2010,Navon:2011,Yefsah:2011,Ku:2012,Mordini:2020}, but these systems are not naturally suited for the textbook gas-expansion experiments.}\BibitemShut {Stop}%
\bibitem [{\citenamefont {Gerbier}\ \emph {et~al.}(2004)\citenamefont {Gerbier}, \citenamefont {Thywissen}, \citenamefont {Richard}, \citenamefont {Hugbart}, \citenamefont {Bouyer},\ and\ \citenamefont {Aspect}}]{Gerbier:2004}%
  \BibitemOpen
  \bibfield  {author} {\bibinfo {author} {\bibfnamefont {F.}~\bibnamefont {Gerbier}}, \bibinfo {author} {\bibfnamefont {J.~H.}\ \bibnamefont {Thywissen}}, \bibinfo {author} {\bibfnamefont {S.}~\bibnamefont {Richard}}, \bibinfo {author} {\bibfnamefont {M.}~\bibnamefont {Hugbart}}, \bibinfo {author} {\bibfnamefont {P.}~\bibnamefont {Bouyer}},\ and\ \bibinfo {author} {\bibfnamefont {A.}~\bibnamefont {Aspect}},\ }\bibfield  {title} {\bibinfo {title} {{Experimental study of the thermodynamics of an interacting trapped Bose-Einstein condensed gas}},\ }\href {https://doi.org/10.1103/PhysRevA.70.013607} {\bibfield  {journal} {\bibinfo  {journal} {Phys. Rev. A}\ }\textbf {\bibinfo {volume} {70}},\ \bibinfo {pages} {013607} (\bibinfo {year} {2004})}\BibitemShut {NoStop}%
\bibitem [{\citenamefont {Nascimb\`ene}\ \emph {et~al.}(2010)\citenamefont {Nascimb\`ene}, \citenamefont {Navon}, \citenamefont {Jiang}, \citenamefont {Chevy},\ and\ \citenamefont {Salomon}}]{Nascimbene:2010}%
  \BibitemOpen
  \bibfield  {author} {\bibinfo {author} {\bibfnamefont {S.}~\bibnamefont {Nascimb\`ene}}, \bibinfo {author} {\bibfnamefont {N.}~\bibnamefont {Navon}}, \bibinfo {author} {\bibfnamefont {K.~J.}\ \bibnamefont {Jiang}}, \bibinfo {author} {\bibfnamefont {F.}~\bibnamefont {Chevy}},\ and\ \bibinfo {author} {\bibfnamefont {C.}~\bibnamefont {Salomon}},\ }\bibfield  {title} {\bibinfo {title} {{Exploring the thermodynamics of a universal Fermi gas}},\ }\href {https://doi.org/10.1038/nature08814} {\bibfield  {journal} {\bibinfo  {journal} {Nature}\ }\textbf {\bibinfo {volume} {{463}}},\ \bibinfo {pages} {1057} (\bibinfo {year} {{2010}})}\BibitemShut {NoStop}%
\bibitem [{\citenamefont {Navon}\ \emph {et~al.}(2011)\citenamefont {Navon}, \citenamefont {Piatecki}, \citenamefont {G\"unter}, \citenamefont {Rem}, \citenamefont {Nguyen}, \citenamefont {Chevy}, \citenamefont {Krauth},\ and\ \citenamefont {Salomon}}]{Navon:2011}%
  \BibitemOpen
  \bibfield  {author} {\bibinfo {author} {\bibfnamefont {N.}~\bibnamefont {Navon}}, \bibinfo {author} {\bibfnamefont {S.}~\bibnamefont {Piatecki}}, \bibinfo {author} {\bibfnamefont {K.}~\bibnamefont {G\"unter}}, \bibinfo {author} {\bibfnamefont {B.}~\bibnamefont {Rem}}, \bibinfo {author} {\bibfnamefont {T.~C.}\ \bibnamefont {Nguyen}}, \bibinfo {author} {\bibfnamefont {F.}~\bibnamefont {Chevy}}, \bibinfo {author} {\bibfnamefont {W.}~\bibnamefont {Krauth}},\ and\ \bibinfo {author} {\bibfnamefont {C.}~\bibnamefont {Salomon}},\ }\bibfield  {title} {\bibinfo {title} {{Dynamics and Thermodynamics of the Low-Temperature Strongly Interacting Bose Gas}},\ }\href {https://doi.org/10.1103/PhysRevLett.107.135301} {\bibfield  {journal} {\bibinfo  {journal} {Phys. Rev. Lett.}\ }\textbf {\bibinfo {volume} {107}},\ \bibinfo {pages} {135301} (\bibinfo {year} {2011})}\BibitemShut {NoStop}%
\bibitem [{\citenamefont {Yefsah}\ \emph {et~al.}(2011)\citenamefont {Yefsah}, \citenamefont {Desbuquois}, \citenamefont {Chomaz}, \citenamefont {G\"unter},\ and\ \citenamefont {Dalibard}}]{Yefsah:2011}%
  \BibitemOpen
  \bibfield  {author} {\bibinfo {author} {\bibfnamefont {T.}~\bibnamefont {Yefsah}}, \bibinfo {author} {\bibfnamefont {R.}~\bibnamefont {Desbuquois}}, \bibinfo {author} {\bibfnamefont {L.}~\bibnamefont {Chomaz}}, \bibinfo {author} {\bibfnamefont {K.~J.}\ \bibnamefont {G\"unter}},\ and\ \bibinfo {author} {\bibfnamefont {J.}~\bibnamefont {Dalibard}},\ }\bibfield  {title} {\bibinfo {title} {{Exploring the Thermodynamics of a Two-Dimensional Bose Gas}},\ }\href {https://doi.org/10.1103/PhysRevLett.107.130401} {\bibfield  {journal} {\bibinfo  {journal} {Phys. Rev. Lett.}\ }\textbf {\bibinfo {volume} {107}},\ \bibinfo {pages} {130401} (\bibinfo {year} {2011})}\BibitemShut {NoStop}%
\bibitem [{\citenamefont {Ku}\ \emph {et~al.}(2012)\citenamefont {Ku}, \citenamefont {Sommer}, \citenamefont {Cheuk},\ and\ \citenamefont {Zwierlein}}]{Ku:2012}%
  \BibitemOpen
  \bibfield  {author} {\bibinfo {author} {\bibfnamefont {M.~J.~H.}\ \bibnamefont {Ku}}, \bibinfo {author} {\bibfnamefont {A.~T.}\ \bibnamefont {Sommer}}, \bibinfo {author} {\bibfnamefont {L.~W.}\ \bibnamefont {Cheuk}},\ and\ \bibinfo {author} {\bibfnamefont {M.~W.}\ \bibnamefont {Zwierlein}},\ }\bibfield  {title} {\bibinfo {title} {{Revealing the Superfluid Lambda Transition in the Universal Thermodynamics of a Unitary Fermi Gas}},\ }\href {https://doi.org/10.1126/science.1214987} {\bibfield  {journal} {\bibinfo  {journal} {Science}\ }\textbf {\bibinfo {volume} {335}},\ \bibinfo {pages} {563} (\bibinfo {year} {2012})}\BibitemShut {NoStop}%
\bibitem [{\citenamefont {Mordini}\ \emph {et~al.}(2020)\citenamefont {Mordini}, \citenamefont {Trypogeorgos}, \citenamefont {Farolfi}, \citenamefont {Wolswijk}, \citenamefont {Stringari}, \citenamefont {Lamporesi},\ and\ \citenamefont {Ferrari}}]{Mordini:2020}%
  \BibitemOpen
  \bibfield  {author} {\bibinfo {author} {\bibfnamefont {C.}~\bibnamefont {Mordini}}, \bibinfo {author} {\bibfnamefont {D.}~\bibnamefont {Trypogeorgos}}, \bibinfo {author} {\bibfnamefont {A.}~\bibnamefont {Farolfi}}, \bibinfo {author} {\bibfnamefont {L.}~\bibnamefont {Wolswijk}}, \bibinfo {author} {\bibfnamefont {S.}~\bibnamefont {Stringari}}, \bibinfo {author} {\bibfnamefont {G.}~\bibnamefont {Lamporesi}},\ and\ \bibinfo {author} {\bibfnamefont {G.}~\bibnamefont {Ferrari}},\ }\bibfield  {title} {\bibinfo {title} {{Measurement of the Canonical Equation of State of a Weakly Interacting 3D Bose Gas}},\ }\href {https://doi.org/10.1103/PhysRevLett.125.150404} {\bibfield  {journal} {\bibinfo  {journal} {Phys. Rev. Lett.}\ }\textbf {\bibinfo {volume} {125}},\ \bibinfo {pages} {150404} (\bibinfo {year} {2020})}\BibitemShut {NoStop}%
\bibitem [{\citenamefont {Etrych}\ \emph {et~al.}(2023)\citenamefont {Etrych}, \citenamefont {Martirosyan}, \citenamefont {Cao}, \citenamefont {Glidden}, \citenamefont {Dogra}, \citenamefont {Hutson}, \citenamefont {Hadzibabic},\ and\ \citenamefont {Eigen}}]{Etrych:2023}%
  \BibitemOpen
  \bibfield  {author} {\bibinfo {author} {\bibfnamefont {J.}~\bibnamefont {Etrych}}, \bibinfo {author} {\bibfnamefont {G.}~\bibnamefont {Martirosyan}}, \bibinfo {author} {\bibfnamefont {A.}~\bibnamefont {Cao}}, \bibinfo {author} {\bibfnamefont {J.~A.~P.}\ \bibnamefont {Glidden}}, \bibinfo {author} {\bibfnamefont {L.~H.}\ \bibnamefont {Dogra}}, \bibinfo {author} {\bibfnamefont {J.~M.}\ \bibnamefont {Hutson}}, \bibinfo {author} {\bibfnamefont {Z.}~\bibnamefont {Hadzibabic}},\ and\ \bibinfo {author} {\bibfnamefont {C.}~\bibnamefont {Eigen}},\ }\bibfield  {title} {\bibinfo {title} {{Pinpointing Feshbach resonances and testing Efimov universalities in \textsuperscript{39}K}},\ }\href {https://doi.org/10.1103/PhysRevResearch.5.013174} {\bibfield  {journal} {\bibinfo  {journal} {Phys. Rev. Res.}\ }\textbf {\bibinfo {volume} {5}},\ \bibinfo {pages} {013174} (\bibinfo {year} {2023})}\BibitemShut {NoStop}%
\bibitem [{Sup()}]{Supplementary}%
  \BibitemOpen
  \href@noop {} {}\bibinfo {note} {See Supplemental Material for details of our numerical calculations and temperature measurements.}\BibitemShut {Stop}%
\bibitem [{\citenamefont {Pethick}\ and\ \citenamefont {Smith}(2002)}]{Pethick:2002}%
  \BibitemOpen
  \bibfield  {author} {\bibinfo {author} {\bibfnamefont {C.}~\bibnamefont {Pethick}}\ and\ \bibinfo {author} {\bibfnamefont {H.}~\bibnamefont {Smith}},\ }\href@noop {} {\emph {\bibinfo {title} {{B}ose--{E}instein Condensation in Dilute Gases}}}\ (\bibinfo  {publisher} {Cambridge University Press},\ \bibinfo {year} {2002})\BibitemShut {NoStop}%
\bibitem [{\citenamefont {Gauthier}\ \emph {et~al.}(2021)\citenamefont {Gauthier}, \citenamefont {Bell}, \citenamefont {Stilgoe}, \citenamefont {Baker}, \citenamefont {Rubinsztein-Dunlop},\ and\ \citenamefont {Neely}}]{Gauthier:2021}%
  \BibitemOpen
  \bibfield  {author} {\bibinfo {author} {\bibfnamefont {G.}~\bibnamefont {Gauthier}}, \bibinfo {author} {\bibfnamefont {T.~A.}\ \bibnamefont {Bell}}, \bibinfo {author} {\bibfnamefont {A.~B.}\ \bibnamefont {Stilgoe}}, \bibinfo {author} {\bibfnamefont {M.}~\bibnamefont {Baker}}, \bibinfo {author} {\bibfnamefont {H.}~\bibnamefont {Rubinsztein-Dunlop}},\ and\ \bibinfo {author} {\bibfnamefont {T.~W.}\ \bibnamefont {Neely}},\ }\bibfield  {title} {\bibinfo {title} {{Dynamic high-resolution optical trapping of atoms}},\ }in\ \href {https://doi.org/10.1016/bs.aamop.2021.04.001} {\emph {\bibinfo {booktitle} {{Advances in Atomic, Molecular and Optical Physics}}}},\ Vol.~\bibinfo {volume} {70},\ \bibinfo {editor} {edited by\ \bibinfo {editor} {\bibfnamefont {L.~F.}\ \bibnamefont {Dimauro}}, \bibinfo {editor} {\bibfnamefont {H.}~\bibnamefont {Perrin}},\ and\ \bibinfo {editor} {\bibfnamefont {S.~F.}\ \bibnamefont {Yelin}}}\ (\bibinfo  {publisher} {Academic Press},\ \bibinfo {year} {2021})\ pp.\ \bibinfo {pages}
  {1--101}\BibitemShut {NoStop}%
\bibitem [{\citenamefont {Myers}\ \emph {et~al.}(2022)\citenamefont {Myers}, \citenamefont {Abah},\ and\ \citenamefont {Deffner}}]{Myers:2022}%
  \BibitemOpen
  \bibfield  {author} {\bibinfo {author} {\bibfnamefont {N.~M.}\ \bibnamefont {Myers}}, \bibinfo {author} {\bibfnamefont {O.}~\bibnamefont {Abah}},\ and\ \bibinfo {author} {\bibfnamefont {S.}~\bibnamefont {Deffner}},\ }\bibfield  {title} {\bibinfo {title} {{Quantum thermodynamic devices: From theoretical proposals to experimental reality}},\ }\href {https://doi.org/10.1116/5.0083192} {\bibfield  {journal} {\bibinfo  {journal} {AVS Quantum Sci.}\ }\textbf {\bibinfo {volume} {4}},\ \bibinfo {pages} {027101} (\bibinfo {year} {2022})}\BibitemShut {NoStop}%
\bibitem [{\citenamefont {Koch}\ \emph {et~al.}(2023)\citenamefont {Koch}, \citenamefont {Menon}, \citenamefont {Cuestas}, \citenamefont {Barbosa}, \citenamefont {Lutz}, \citenamefont {Fogarty}, \citenamefont {Busch},\ and\ \citenamefont {Widera}}]{Koch:2023}%
  \BibitemOpen
  \bibfield  {author} {\bibinfo {author} {\bibfnamefont {J.}~\bibnamefont {Koch}}, \bibinfo {author} {\bibfnamefont {K.}~\bibnamefont {Menon}}, \bibinfo {author} {\bibfnamefont {E.}~\bibnamefont {Cuestas}}, \bibinfo {author} {\bibfnamefont {S.}~\bibnamefont {Barbosa}}, \bibinfo {author} {\bibfnamefont {E.}~\bibnamefont {Lutz}}, \bibinfo {author} {\bibfnamefont {T.}~\bibnamefont {Fogarty}}, \bibinfo {author} {\bibfnamefont {T.}~\bibnamefont {Busch}},\ and\ \bibinfo {author} {\bibfnamefont {A.}~\bibnamefont {Widera}},\ }\bibfield  {title} {\bibinfo {title} {{A quantum engine in the BEC–BCS crossover}},\ }\href {https://doi.org/10.1038/s41586-023-06469-8} {\bibfield  {journal} {\bibinfo  {journal} {Nature}\ }\textbf {\bibinfo {volume} {621}},\ \bibinfo {pages} {723} (\bibinfo {year} {2023})}\BibitemShut {NoStop}%
\bibitem [{\citenamefont {Estrada}\ \emph {et~al.}(2024)\citenamefont {Estrada}, \citenamefont {Mayo}, \citenamefont {Roncaglia},\ and\ \citenamefont {Mininni}}]{Estrada:2024}%
  \BibitemOpen
  \bibfield  {author} {\bibinfo {author} {\bibfnamefont {J.~A.}\ \bibnamefont {Estrada}}, \bibinfo {author} {\bibfnamefont {F.}~\bibnamefont {Mayo}}, \bibinfo {author} {\bibfnamefont {A.~J.}\ \bibnamefont {Roncaglia}},\ and\ \bibinfo {author} {\bibfnamefont {P.~D.}\ \bibnamefont {Mininni}},\ }\bibfield  {title} {\bibinfo {title} {{Quantum engines with interacting Bose-Einstein condensates}},\ }\href {https://doi.org/10.1103/PhysRevA.109.012202} {\bibfield  {journal} {\bibinfo  {journal} {Phys. Rev. A}\ }\textbf {\bibinfo {volume} {109}},\ \bibinfo {pages} {012202} (\bibinfo {year} {2024})}\BibitemShut {NoStop}%
  \bibitem [{rep()}]{repository}%
  \BibitemOpen
  \href@noop {} {}\bibinfo {note} {{\href{https://doi.org/10.17863/CAM.116975}{https://doi.org/10.17863/CAM.116975}}}\BibitemShut {NoStop}%
\bibitem [{\citenamefont {Eigen}\ \emph {et~al.}(2016)\citenamefont {Eigen}, \citenamefont {Gaunt}, \citenamefont {Suleymanzade}, \citenamefont {Navon}, \citenamefont {Hadzibabic},\ and\ \citenamefont {Smith}}]{Eigen:2016}%
  \BibitemOpen
  \bibfield  {author} {\bibinfo {author} {\bibfnamefont {C.}~\bibnamefont {Eigen}}, \bibinfo {author} {\bibfnamefont {A.~L.}\ \bibnamefont {Gaunt}}, \bibinfo {author} {\bibfnamefont {A.}~\bibnamefont {Suleymanzade}}, \bibinfo {author} {\bibfnamefont {N.}~\bibnamefont {Navon}}, \bibinfo {author} {\bibfnamefont {Z.}~\bibnamefont {Hadzibabic}},\ and\ \bibinfo {author} {\bibfnamefont {R.~P.}\ \bibnamefont {Smith}},\ }\bibfield  {title} {\bibinfo {title} {{Observation of Weak Collapse in a Bose-Einstein Condensate}},\ }\href {https://doi.org/10.1103/PhysRevX.6.041058} {\bibfield  {journal} {\bibinfo  {journal} {Phys. Rev. X}\ }\textbf {\bibinfo {volume} {6}},\ \bibinfo {pages} {041058} (\bibinfo {year} {2016})}\BibitemShut {Stop}%
\end{thebibliography}
\end{document}